\documentclass[prd,aps,twocolumn,preprintnumbers,showpacs, nofootinbib,superscriptaddress,notitlepage]{revtex4-1}
\usepackage{amsmath,graphicx,epsfig,amssymb}
\usepackage{inputenc}
\usepackage[usenames]{color}
\usepackage{slashed}
\usepackage{multirow}
\usepackage{subfigure}
\usepackage{dsfont}

\allowdisplaybreaks

\usepackage{floatrow}
\newfloatcommand{capbtabbox}{table}[][\FBwidth]

\begin{document}

\title{Conformal Mapping in Matching Quark Correlation Functions to Parton Distribution Functions}

\author{Jia-Lu Zhang}\email{Corresponding author: elpsycongr00@sjtu.edu.cn }
\affiliation{INPAC, Key Laboratory for Particle Astrophysics and Cosmology (MOE),  Shanghai Key Laboratory for Particle Physics and Cosmology, School of Physics and Astronomy, Shanghai Jiao Tong University, Shanghai 200240, China}
\begin{abstract}
In high-energy particle physics, extracting parton distribution functions (PDFs) from lattice quantum chromodynamics (QCD) calculations remains a significant challenge, particularly due to the divergent nature of perturbative expansions at high orders. The presence of renormalon singularities in the Borel plane further hinders the accurate determination of PDFs, especially in the context of lattice QCD and effective field theory approaches like Large Momentum Effective Theory (LaMET). This study explores the application of conformal mapping as a technique to improve the convergence of perturbative series for matching kernels. By transforming the Borel plane to map singularities onto the unit disk, this method mitigates the effects of divergent behavior in high-order perturbative expansions of matching kernels. The numerical analysis focuses on the matching kernel for quark correlation functions (QCFs), using the CT18NNLO PDFs for u-quarks and d-quarks, along with N3LO hard kernel inputs. The results demonstrate that conformal mapping enhances the stability of the perturbative series, reducing the root-mean-square (RMS) error by up to $40\%$ compared to conventional $\alpha_s$ series. These findings highlight the potential of conformal mapping to enhance the precision of PDFs and reduce theoretical uncertainties in high-order QCD calculations.
\end{abstract}

\maketitle
\section{introduction}
PDFs are fundamental quantities in QCD that describe the momentum distribution of quarks and gluons inside a hadron. PDFs play a critical role in high-energy physics, providing essential input for the computation of cross-sections in processes involving hadrons at particle colliders, such as the Large Hadron Collider (LHC). Moreover, PDFs serve as a vital tool for testing the Standard Model and probing potential avenues for new physics~\cite{Alekhin:2000ch,Hou:2019efy,Brivio:2017vri,Beacham:2019nyx}. Consequently, the precise determination of PDFs is of paramount importance. However, their extraction is inherently challenging due to the non-perturbative nature of QCD at low energy scales. Recent advancements in lattice QCD, particularly those employing factorization techniques~\cite{Cichy:2018mum, Zhao:2018fyu, Ji:2020ect, Izubuchi:2018srq, Radyushkin:1998es, Hua:2020gnw, Radyushkin:2017cyf,Ji:2014gla,Constantinou:2022yye} inspired by LaMET~\cite{Ji:2013dva}, have opened new avenues for extracting PDFs from lattice calculations.

Factorization techniques in this context typically involve computing space-like correlators using lattice QCD at finite momentum and performing a matching procedure to connect these quantities with the physical PDFs defined in the continuum limit. An extraction of PDF with high precision typically relies on a high-loop calculation of matching kernel, which requires calculating terms up to next-to-next-to-leading order (NNLO) or even next-to-next-to-next-to-leading order (N3LO) to achieve the desired precision. However, QCD perturbative expansions, particularly at high orders, often display asymptotic, divergent behavior due to  singularities in the Borel plane~\cite{Dyson:1952tj,Beneke:1994qe}. These singularities arise from contributions of instantons~\cite{Belavin:1975fg,Brezin:1977gk} and renormalons~\cite{Gross:1974jv,Beneke:1998ui,Beneke:1994sw,Beneke:1992ch}. This divergence poses a significant challenge, as it limits the accuracy and reliability of the extracted PDFs.

To address the issue, conformal mapping has emerged as a powerful tool for improving the series' behavior~\cite{Frazer:1961zz,Seznec:1979ev,Caprini:2021wvf,Caprini:2024unl}. Rooted in complex analysis, the conformal mapping technique works by transforming the Borel plane, where the series is typically asymptotic, into the plane of a new variable. This transformation maps the cut Borel plane onto the unit disk, redistributing the singularities in a way that accelerates the convergence of the series and expands its domain of convergence. As a result, the conformal-mapped series exhibits a "tamed" large-order behavior, becoming less sensitive to the factorial growth of renormalons that drive the divergence of perturbative expansions in QCD.

Moreover, conformal mapping in the Borel plane accelerates the series, enabling more accurate determinations of PDFs without increasing the level of loop corrections. This method also reduces the sensitivity of the matching kernel to factorization scale choices, thereby further improving the precision of the calculation.

In this paper, I demonstrate how conformal mapping significantly improves the application of factorization theorems by enhancing the convergence of Borel-transformed expansions of matching kernels.  This not only stabilizes the perturbative series but also reduces the theoretical uncertainties in the matching procedure, making it a crucial tool for achieving precision at higher orders.

The remainder of this paper is organized as follows. In Section II, I provide a brief review of the process for matching QCFs to PDFs. Section III introduces the fundamentals of conformal mapping and discusses its application to the QCF-to-PDF matching process. Section IV presents the numerical results, demonstrating the improved convergence of the enhanced series. Finally, I conclude with a summary in Section V.
\section{Matching procedure}
The PDFs of a proton are defined by
\begin{equation}
    \begin{aligned}
\label{PDF}
 f_q(x,\mu^2) =&\int_{-\infty}^{\infty}\frac{dz}{4\pi}e^{-ixz P^+}\\
 &\times\langle P|\bar{\psi}_q(z n_+)n\!\!\!\slash_+W_c(z n_+,0)\psi_q(0)|P\rangle,    
    \end{aligned}
\end{equation}  
where $n_+^\mu= (1,0,0,-1)/\sqrt{2}$ is a unit lightcone vector. $\psi_q(x)$ stands for a light quark with flavour $q$. The momentum of the proton is directed along the $n_-^\mu=(1,0,0,-1)/\sqrt{2}$, expressed as $P^\mu=P^{+}  n_-^\mu = (P^z,0,0,P^z)$.
The Wilson line $W_c(x,y)$, which is introduced to maintain gauge invariance, is given explicitly by
\begin{equation}
W_c(x, y)=\mathcal{P} \exp \left[i g \int_0^1 \mathrm{~d} t(x-y)_\mu A^\mu(t x+(1-t) y)\right].
\end{equation}

Quark correlation functions(QCFs) are defined as
 \begin{align}
\label{QCF_def}
 &\tilde f^\nu_q(\omega,\xi^2,\mu^2)= \langle P|\bar{\psi}_q^b(\xi)\gamma^\nu W_c(\xi,0)\psi_q^b(0)| P\rangle
\end{align}   
where $\xi$ is a spacelike vector and $\omega=P\cdot \xi$.

To connect QCFs and PDFs, one utilizes a factorization procedure~\cite{Radyushkin:2017cyf}  
\begin{eqnarray} 
\label{matching}
 \tilde f^{\nu}_R(\omega,\xi^2,\mu^2)=\int_{-1}^1\frac{dx}xf(x,\mu^2)K^\nu\left(x\omega,\xi^2,\mu^2\right), 
\end{eqnarray}
where the subscript $R$ denotes renormalized quantities. The factorization ensures that the short-distance physics, encoded in the matching kernel, is separated from the long-distance hadronic structure, represented by the PDF. The matching kernel is  perturbatively calculable, and has many numerical results in either coordinate space and momentum space~\cite{Izubuchi:2018srq,Chen:2020arf,Chen:2020iqi,Chen:2020ody}. Recently, it has been determined to three-loop accuracy in $\overline{\mathrm{MS}}$ scheme in recent calculations~\cite{Cheng:2024wyu}, which is used as the input of this paper. 

However, the overarching goal is to utilize this matching relation to extract PDFs from current lattice QCD data~\cite{LatticeParton:2022zqc,LatticePartonCollaborationLPC:2022myp,LatticeParton:2018gjr}. In lattice QCD, non-local operators containing Wilson lines are subject to linear divergences, necessitating additional renormalization procedures, such as the ratio scheme and the hybrid renormalization scheme~\cite{Ji:2020brr,Han:2024ucv,Han:2023xbl,Han:2023hgy,Liu:2018tox,Gao:2021dbh,Chou:2022drv,Su:2022fiu,Ji:2022ezo,Gao:2022ytj}. Consequently, N3LO matching kernels in these lattice-preferred renormalization schemes are highly anticipated.
\section{conformal mapping}
\subsection{Theoretical Foundation}

In QCD, it is well-known that the perturbative expansion in the strong coupling constant $\alpha_s$ is asymptotic. This implies that the series has zero radius of convergence with respect to $\alpha_s$ and diverges as the order of $\alpha_s$ increases. As a result, summing the series directly does not provide an accurate value for physical quantities.

One effective way to extract a meaningful result from such an asymptotic series is through the Borel summation technique. Consider the asymptotic series for a physical observable 
$\phi(\alpha_s)$, written in terms of powers of $\alpha_s$ as

\begin{equation}
\label{series}
    \phi(\alpha_s) = \sum_{i=0}^{\infty} c_i \, \alpha_s^{i+1},
\end{equation}
where the coefficients \( c_i \) are known. The first step in the Borel summation procedure is to apply the Borel transform to the series, defined as

\begin{equation}
    \mathcal{B}[\phi](u) = \sum_{k=0}^{\infty} \frac{c_k}{k!} \left( \frac{u}{\beta_0} \right)^k,
\end{equation}
which transforms the series into a new function of the variable $u$. The next step involves calculating the Borel integral, which is given by

\begin{equation}
    \Phi(\alpha_s) = \frac{1}{\beta_0} \int_0^{\infty} \mathrm{d}u \, e^{-w / (\beta_0 \alpha_s)} \, \mathcal{B}[\phi](u),
\end{equation}
where \( \beta_0 =11-\frac{3}{2}n_f \) is the first coefficient of the $\beta$ function. If the integral converges, the function \(\Phi(\alpha_s) \) will share the same asymptotic expansion as the original series \(\phi(\alpha_s)\), but it can provide a more precise result by overcoming the limitations of the divergent perturbative series.

However, in many cases, the Borel integral is ill-defined due to the presence of singularities along the integration path, typically caused by branch points or poles in the Borel plane. To address this issue, the conformal mapping technique is often used to transform the Borel plane into a region where the integral becomes well-defined and convergent.

This method involves reparametrizing the variable $u$ through a conformal mapping that deforms the Borel plane, such that singularities,  are mapped onto the boundary of a unit disk. This approach enlarges the domain of convergence and improves the asymptotic convergence rate of perturbative expansions, allowing for better analytic continuation and control of divergences. The conformal mapping is achieved through the following transformation:

\begin{equation}
    \tilde{w}(u) = \frac{\sqrt{1 - u / u_{\mathrm{UV}}} - \sqrt{1 - u / u_{\mathrm{IR}}}}{\sqrt{1 - u / u_{\mathrm{UV}}} + \sqrt{1 - u / u_{\mathrm{IR}}}},
\end{equation}
where \( u_{\mathrm{UV}} \) and \( u_{\mathrm{IR}} \) denote the positions of the nearest singularities on the negative and positive real axes, respectively, which are mapped onto the unit disk in the transformed plane. 

With this transformation, the Borel transformed series can now be expressed in terms of the new variable $w=\tilde{w}(u)$ as

\begin{equation}
    \mathcal{B}[\phi](w) = \sum_{k=0}^{\infty} c'_k\left( \frac{w}{\beta_0} \right)^k,
\end{equation}
where the coefficients $c'_k$ can be computed through an order-by-order iteration.

The original series can then be re-expressed in terms of the transformed coefficients and the new integration variables

\begin{equation}
\label{npseries}
    \phi(\alpha_s) = \sum_{i=0}^{\infty} c'_i \, \mathcal{W}_i(\alpha_s),
\end{equation}
where \( \mathcal{W}_i(\alpha_s) \) is defined as

\begin{equation}
\label{W}
    \mathcal{W}_i(\alpha_s) = \frac{1}{\beta_0} \, \mathrm{PV} \int_0^\infty e^{-u / (\beta_0 \alpha_s)} \, \tilde{w}(u)^i \, \mathrm{d}u.
\end{equation}
Here, \( \mathrm{PV} \) denotes the principal value of the integral.

Through conformal mapping, the cuts along the real axis is transformed onto the unit circle, $|w|=1$, with the first Riemann sheet of the holomorphic domain mapped onto the interior of the unit disk. Consequently, the new series in Eq.~(\ref{npseries}) exhibits convergence throughout the entire u-plane, except along the cuts on the real axis.

It is important to recognize that the renormalon singularities on the positive real axis underline the necessity of including non-perturbative terms in the perturbative series, which are associated with higher-twist operators. While conformal mapping aids in improving convergence, it does not eliminate the requirement for these non-perturbative contributions. To incorporate them, one should use resurgence theory~\cite{Maiezza:2021mry,Aniceto:2018bis}, which connects these additional terms to the computation of the discontinuity in Eq.~(\ref{W})~\cite{Caprini:2024unl}.

The series in Eq.~(\ref{npseries}) are commonly referred to as non-power series. In the remainder of the paper, I will refer to the series in Eq.~(\ref{series}) as $\alpha_s$ series.

\subsection{Conformal mapping of QCF}
To apply conformal mapping to the matching kernel in Eq.~(\ref{matching}), one should re-express it as the series in $\alpha_s$:

\begin{equation}
 K^\nu\left(x\omega,\xi^2,\mu^2\right)=\sum_{i=0}^\infty \alpha_s(\mu)^i*K^\nu_i\left(x\omega,\xi^2,\mu^2\right),
\end{equation}
where $K^\nu_n\left(x\omega,\xi^2,\mu^2\right)$ denotes the hard kernel at $n$ loop.
Then the factorization in Eq.~(\ref{matching}) is written as
\begin{equation}
\tilde f^{\nu}_q(\omega,\xi^2,\mu^2)=\sum_{i=0}^\infty c^\nu_i(\omega,\xi^2,\mu^2)\alpha_s(\mu)^i, \end{equation}
where
\begin{equation}
c^\nu_i(\omega,\xi^2,\mu^2)=\int_{-1}^1\frac{dx}{x}f_q(x,\mu^2)K^\nu_i\left(x\omega,\xi^2,\mu^2\right). 
\end{equation}

The next step is determine  the nearest singularities of $\mathcal{B}[\tilde{f}^\nu_q](u)$, which is the Borel transformation of $\tilde{f}^\nu_q(\omega,\xi^2,\mu^2)$. The nearest singularity of $\mathcal{B}[\tilde{f}^\nu_q](u)$ is at $u_{\mathrm{UV}}=-1$ and $u_{\mathrm{IR}}=\frac{1}{2}$. It is crucial to highlight that $u_{\mathrm{IR}}$, contrary to what its name might imply, corresponds to an UV singularity rather than an infrared IR one. This distinction is significant, as the UV nature of this renormalon originates from the spacelike Wilson line that appears in the definition of QCFs. Such UV renormalons have been a topic of extensive study, particularly in relation to quasi-PDFs~\cite{Liu:2023onm,Holligan:2023rex} and within the framework of HQET~\cite{Beneke:1998ui,Beneke:1994sw, Han:2024cht}. 

It has been widely demonstrated~\cite{Ji:2020brr,Orginos:2017kos} that, within the ratio scheme or hybrid renormalization scheme, the UV renormalon at \( u = \frac{1}{2} \) is effectively canceled. As a result, the singularity structure shifts to \( u_{\mathrm{UV}} = -1 \) and \( u_{\mathrm{IR}} = 2 \). Consequently, the matching kernel in these schemes exhibits potentially better convergence behavior compared to that in the \( \overline{\mathrm{MS}} \) scheme. This observation underscores the importance of developing high-order matching kernels within the ratio scheme or hybrid renormalization scheme.

\begin{figure*}[ht]
\centering
\begin{minipage}[t]{0.48\textwidth}
    \centering  \includegraphics[width=\textwidth]{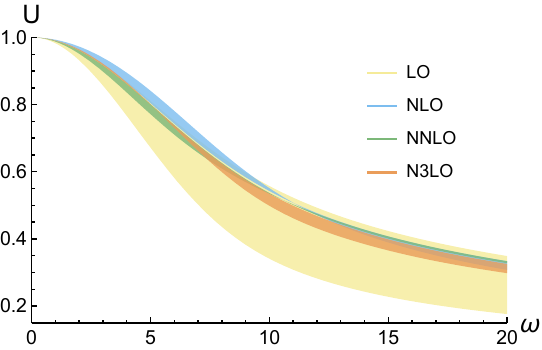}
    \centerline{(a)}
\end{minipage}
\hfill
\begin{minipage}[t]{0.48\textwidth}
    \centering
\includegraphics[width=\textwidth]{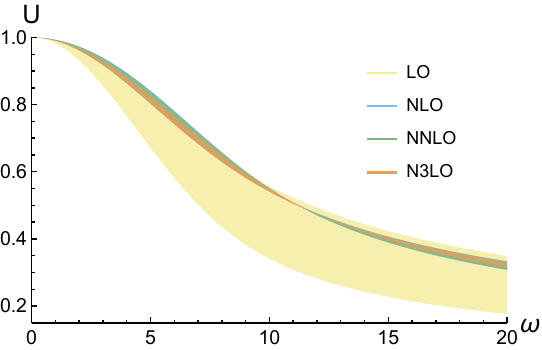}
    \centerline{(b)}
\end{minipage}
\begin{minipage}[t]{0.48\textwidth}
    \centering    \includegraphics[width=\textwidth]{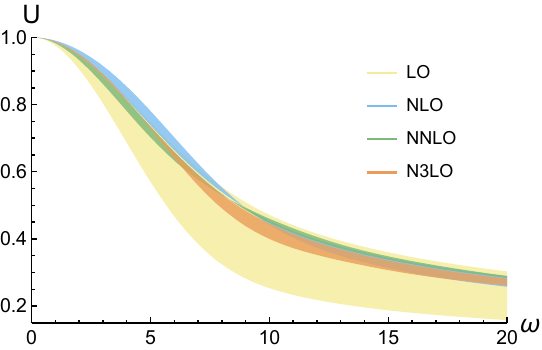}
    \vskip 1ex 
    (c)
\end{minipage}
\hfill
\begin{minipage}[t]{0.48\textwidth}
    \centering
\includegraphics[width=\textwidth]{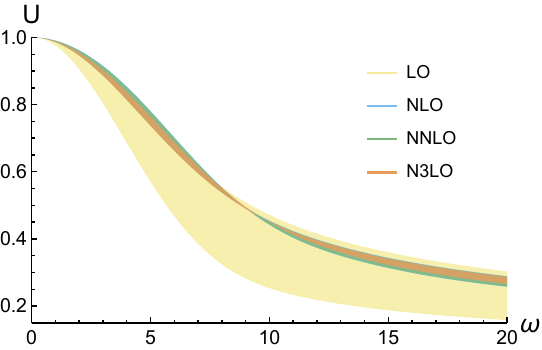}
\vskip 1ex 
(d)

\end{minipage}
\caption{The QCFs \( U(\omega, \xi^2) \) for \( u \)-quark and \( d \)-quark at different orders. The error bands originate from the variation of the factorization scale between \( 1.3~\mathrm{GeV} \) and \( 15~\mathrm{GeV} \), with \( \xi \) fixed at \( |\xi| = 1/2~\mathrm{GeV} \). (a) illustrates the QCFs of the
u-quark in the $\alpha_s$ series, while (b) presents those of the d-quark. Similarly, (c) depicts the QCFs of the 
u-quark in the non-power series, and (d) displays those of the
d-quark.}
\label{QCF}
\end{figure*}
\section{numerical calculation}
In this section, I present a numerical analysis of the impact of conformal mapping on the matching kernel for QCFs. The analysis utilizes the CT18NNLO PDFs for the
u-quark and
d-quark~\cite{Hou:2019efy}, along with the N3LO hard kernel~\cite{Cheng:2024wyu} as inputs. All data are renormalized in the $\overline{\mathrm{MS}}$ scheme.

In FIG.\ref{QCF}, I show the normalized QCF $U\left(\omega,\xi^2\right) = \frac{\mathrm{i}}{4\omega}\xi\cdot F_{u}\left(\omega,\xi^2\right)/\lim\limits_{\omega\to0}\frac{\mathrm{i}}{4\omega}\xi\cdot F_{u}\left(\omega,\xi^2\right)$ at $1/|\xi|= 2~\mathrm{GeV}$, calculated using matching kernels in 
$\alpha_s$ series and non-power series expansions, respectively. The error bands originate from variations in the factorization scale between 1.3 GeV and 15 GeV. A closer examination reveals that the matching kernels in non-power series expansions improve the stability of higher-order QCFs. Specifically, QCFs calculated with matching kernels $\alpha_s$ series show wider error bands, with curves deviating more significantly as the order $\alpha_s$ increases. In contrast, the QCFs in non-power series expansions have narrower error bands, and higher-order results remain closely aligned with lower-order QCFs. This suggests that the non-power series approach effectively mitigates the instabilities typically observed in perturbative series as higher orders are included.

\begin{figure*}[ht]
\centering
\begin{minipage}[t]{0.48\textwidth}
    \centering    \includegraphics[width=\textwidth]{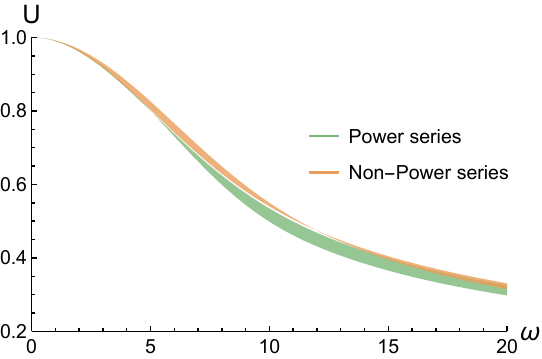}
    \vskip 1ex 
    (a)
\end{minipage}
\hfill
\begin{minipage}[t]{0.48\textwidth}
    \centering
\includegraphics[width=\textwidth]{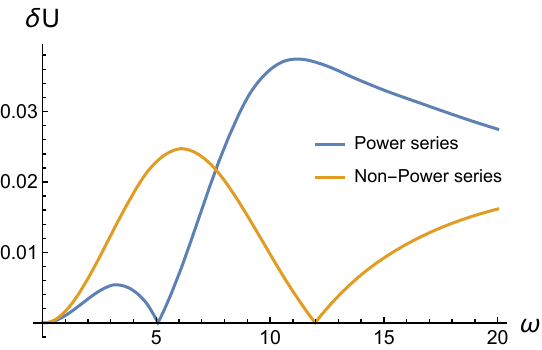}
\vskip 1ex 
(b)
\end{minipage}
\begin{minipage}[t]{0.48\textwidth}
    \centering    \includegraphics[width=\textwidth]{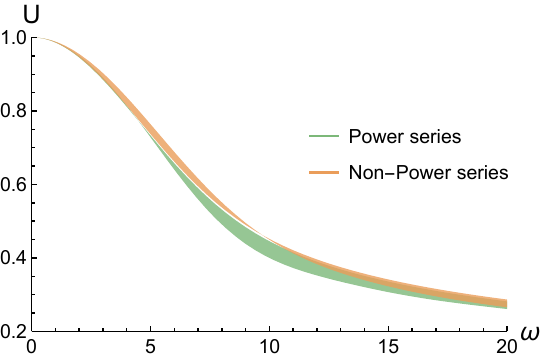}
    \vskip 1ex 
    (c)
\end{minipage}
\hfill
\begin{minipage}[t]{0.48\textwidth}
    \centering
\includegraphics[width=\textwidth]{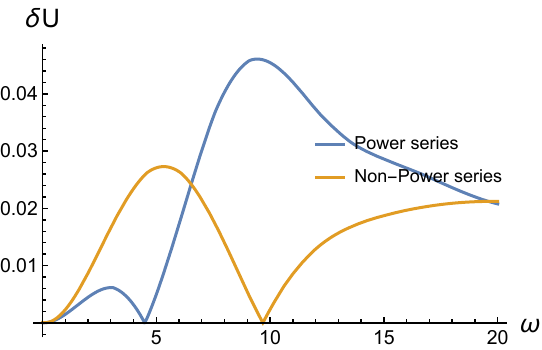}
\vskip 1ex 
(d)
\end{minipage}
\caption{The N3LO QCFs \( U(\omega, \xi^2) \) for \( u \)-quark and \( d \)-quark. The error bands originate form the variation of the factorization scale between \( 1.3~\mathrm{GeV} \) and \( 15~\mathrm{GeV} \), with \( \xi \) fixed at \( |\xi| = 1/2~\mathrm{GeV} \). (a) displays the N3LO QCFs of the \( u \)-quark in both the \( \alpha_s \) series and the non-power series expansion, while (c) presents those of the \( d \)-quark. (b) illustrates the width of the N3LO error band for the \( \alpha_s \) series and the non-power series expansion for the \( u \)-quark, and (d) shows the corresponding width for the \( d \)-quark.}
\label{N3LO}
\end{figure*}

In FIG.~\ref{N3LO}, a comparison of N3LO QCFs in $\alpha_s$ series and non-power series expansions further emphasizes the advantages of conformal mapping. A notable feature is the unique point where the error band width is zero, marking the intersection of curves corresponding to different factorization scales. This phenomenon illustrates that the width of the error band alone cannot fully describe the convergence behavior of the series. A similar trend is reported in FIG.4 of~\cite{Chen:2019lzz}, highlighting a broader pattern across different studies.

It is essential to recognize the limitations of the analysis. The PDFs used are restricted to NNLO, precluding complete cancellation of factorization scale dependence at N3LO in the matching kernel. This residual dependence underscores the necessity of further refinements in PDF calculations to achieve full consistency at higher orders.

However, calculating the root-mean-square (RMS) of the error band 
\begin{equation}
\sigma_{RMS}=\sqrt{\frac1N\sum_{i=1}^N\left(\frac{y_{i,\text{upper}}-y_{i,\text{lower}}}2\right)^2},
\end{equation}
remains a plausible approach to assess which series provides a better approximation.

For the N3LO $\alpha_s$ series expansion of the u-quark, the RMS error is $\sigma_{\mathrm{RMS}}\sim 1.31\times 10^{-2}$, while for the non-power series, it reduces significantly to $\sigma_{\mathrm{RMS}}\sim 7.04\times 10^{-3}$. Similarly, for the d-quark, the RMS error decreases from $\sigma_{\mathrm{RMS}}\sim 1.40\times 10^{-2}$ ($\alpha_s$ series) to $\sigma_{\mathrm{RMS}}\sim 8.72\times 10^{-3}$ (non-power series). These results demonstrate a convergence improvement of approximately $40\%$, affirming the efficacy of conformal mapping in enhancing perturbative stability.

\section{summary and prospect}
This paper discusses the application of conformal mapping to improve the convergence of perturbative series in the extraction of PDFs from lattice QCD calculations. The perturbative expansion in QCD is known to be asymptotic and diverges at high orders due to renormalon singularities in the Borel plane. Conformal mapping addresses this issue by transforming the Borel plane, making most use of the holomorphic domain, and thus improving the convergence of the perturbative series.

The key application in this study involves the matching procedure for QCFs, which connects lattice QCD results to PDFs. By using the conformal mapping technique, the Borel transformed series of matching kernels becomes more stable, leading to better convergence properties. Numerical results using the CT18NNLO PDFs for the u-quark and d-quark, and N3LO hard kernel data show that the conformal mapping approach, significantly reduces the RMS error compared to the traditional $\alpha_s$ series expansion.

The results highlight that conformal mapping enhances the stability and reliability of high-order QCD calculations, thus reducing theoretical uncertainties in PDF extraction. This approach represents a valuable advancement for precision studies in QCD, especially within the lattice QCD framework and effective theory applications. 

Future work should focus on applying this non-power matching kernel with lattice data to achieve more precise PDF extractions. Furthermore, this technique demonstrates potential for application in the extraction of Distribution Amplitudes (DAs) and Generalized Parton Distributions (GPDs). Additionally, I aim to explore opportunities to employ these methods for a more precise determination of DAs within alternative renormalization schemes, such as the ratio scheme and hybrid renormalization scheme.

\section*{Acknowledgement}
I would like to thank Dr. Chao Han and Prof. Wei Wang for their valuable discussions. I am especially grateful to Xiang Li for the detailed explanation of the N3LO matching kernel in the paper~\cite{Cheng:2024wyu}.

\end{document}